\documentclass[12pt]{article}

\usepackage[english]{babel}

\begin{document}

\title{Anomalous properties of the Kronig-Penney model with compositional
and structural disorder}

\author{ J. C. Hern\'{a}ndez Herrej\'{o}n${}^{1}$, F.~M.~Izrailev${}^{2}$, and
L.~Tessieri${}^{1}$ \\
{\it ${}^{1}$ Instituto de F\'{\i}sica y Matem\'{a}ticas} \\
{\it Universidad Michoacana de San Nicol\'{a}s de Hidalgo} \\
{\it 58060, Morelia, Mexico} \\
{\it ${}^{2}$ Instituto de F\'{\i}sica, Universidad Aut\'{o}noma de Puebla,} \\
{\it Puebla, 72570, Mexico}}

\date{14th January 2008}

\maketitle

\begin{abstract}
We study the localization properties of the eigenstates in the
Kronig-Penney model with weak compositional and structural disorder.
The main result is an expression for the localization length
that is
valid for any kind of self- and inter-correlations of the two types of
disorder. We show that the interplay between compositional and
structural disorder can result in anomalous localization.
\end{abstract}

Pacs numbers: 73.20.Jc, 73.20.Fz, 71.23.An
\bigskip

Recently, much attention was paid to low-dimensional disordered
models with long-range correlations in random potentials. Apart from
the theoretical aspects, the interest on this issue has increased
significantly due to the possibility of constructing random
potentials with specific correlations which result in a strong
enhancement or reduction of the localization
length~\cite{Mou98,Izr99,Izr01,Izr05,Kuh00,Kro02}.
These new effects allow for the fabrication of electron and
optic/electromagnetic devices with desired anomalous transport properties.
As was shown analytically~\cite{Izr99,Izr01,Izr05} and confirmed
experimentally~\cite{Kuh00,Kro02}, one can arrange prescribed windows of
energy with perfect transmission (or reflection) of scattering waves.

One of the most important models, both from the theoretical and
experimental point of view, is the Kronig-Penney (KP) model, which was
introduced long ago to analyze electronic states in crystals~\cite{Kro31}.
Since the '80s, this model has attracted considerable attention because
it provides a convenient description of superlattices (see,
e.g.,~\cite{Dav98} and references therein).
Modifications of the standard Kronig-Penney model have been
suggested for a study of the physics of random and quasi-periodic
systems with various applications, see, e.g.,~\cite{Koh83}.
Recently, the Kronig-Penney model has been used to discuss the
possibility of selective transmission in waveguides
(see~\cite{Kuh00,Kro02} and references therein).

In this paper we study the KP model with two types of weak disorder.
Disorder of the first kind, or ``compositional'', is due to small
variations in strength of the delta-shaped barriers. In addition, the
spacings between the barriers can be also randomly perturbed (the
so-called ``structural'' disorder). Our interest lies in the interplay
of these two kinds of disorder which can exhibit both self-correlations
and mutal correlations.
Our goal is to derive a formula for the localization length, and to
analyze it.

The stationary Schr\"{o}dinger equation for the eigenstates $\psi(x)$
has the form
\begin{equation}
-\frac{\hbar^{2}}{2m} \psi^{\prime \prime}(x) +
\sum_{n=-\infty}^{\infty} U_n \delta(x-x_{n}) \psi(x) = E \psi(x),
\label{kpmodel}
\end{equation}
where $U_n=U+u_n$ and $x_n$ are amplitude and position of the
$n$-th $\delta$-barrier.
In what follows we use units in which $\hbar^{2}/2m = 1$; we can thus
write the energy of the eigenstates as $E = q^{2}$ where $q$ is the
electron wavenumber.

The positions of the $\delta$-barriers are assumed to be slightly
shifted with respect to the lattice sites, $x_{n} = na + a_{n}$,
where $a$ is the lattice step. The variables $u_{n}$ represent
fluctuations of the barrier strength around the mean value $U$. Our
analysis is restricted to the case of weak disorder for which both
variables $u_n$ and $a_{n}$ have zero average, $\langle u_n \rangle = 0$
and $\langle a_{n} \rangle = 0$, and small variances,
$q^{2} \langle a_{n}^{2} \rangle \ll 1$ and
$\langle u_{n}^{2} \rangle \ll U^{2}$. We remark that the condition
$q^{2} \langle a_{n}^{2} \rangle \ll 1$ implies that the energy must
be low on a scale set by $1/\langle a_{n}^{2} \rangle$.
Note that in contrast with many previous studies, {\it both}
variables are random and may have stationary correlations: our main
interest lies in how these correlations shape the properties of the
localization length $l_{\rm loc}(E)$ of the eigenstates.

It is convenient to introduce the relative displacements of the
barriers, $\Delta_{n} = a_{n+1} - a_{n}$, having zero mean,
$\langle \Delta_{n} \rangle = 0$, and small variance,
$q^{2} \langle \Delta_{n}^{2} \rangle \ll 1$.
Apart from the first two moments of the random variables $\Delta_{n}$ and
$u_{n}$, one has to give the binary correlators,
\begin{equation}
\begin{array}{ccl}
\chi_{1} (k) & = &
\langle u_{n} u_{n+k} \rangle / \langle u_{n}^{2} \rangle \\
\chi_{2} (k) & = &
\langle \Delta_{n} \Delta_{n+k} \rangle / \langle \Delta_{n}^{2} \rangle \\
\chi_{3} (k) & = & \langle u_{n} \Delta_{n+k} \rangle /
\langle u_{n} \Delta_{n} \rangle .
\end{array}
\label{bincor}
\end{equation}
We will not attribute specific forms to the correlators
$\chi_{i}(k)$; we simply assume that they depend only on the index
difference $k$ because of the spatial homogeneity in the mean of the
model and that they are even functions of $k$.

It is worthwhile to note that Eq.~(\ref{kpmodel}) can be treated as
the wave equation for electromagnetic waves in a one-dimensional (1D)
waveguide with wavenumber $q=\omega/c$. Therefore our results
are equally applicable to the classical scattering in optical and
microwave devices of the Kronig-Penney type with correlated
disorder. Our model is also equivalent to a classical oscillator with
a parametric perturbation constituted by a succession of $\delta$-kicks
whose amplitudes and time-dependence are determined by $U_n$ and $a_n$.
This correspondence allows one to cast Eq.~(\ref{kpmodel}) in the form
\begin{equation}
\ddot{x} + [ q^{2} - \sum_{n=-\infty}^{\infty} U_{n}
\delta \left( t - t_{n} \right) ] x = 0.
\label{dyneq}
\end{equation}
Our analysis is based on the Hamiltonian approach~\cite{Izr95,Izr98}
according to which the spatial structure of eigenstates of the
KP-model can be analyzed by exploring the time evolution of
the kicked oscillator described by the dynamical
equation~(\ref{dyneq}). Such a dynamical approach considers the
Schr\"{o}dinger equation as an initial-value problem and can be
treated as a modification of the transfer matrix approach.

Integrating the dynamical equation~(\ref{dyneq}) between two successive
kicks, one obtains the map
\begin{equation}
\begin{array}{ccl}
x_{n+1} & = & \left[ \left( U_{n}/q \right) \sin \left( \mu + \mu_{n} \right)
+ \cos \left( \mu + \mu_{n} \right) \right] x_{n} \\
& + & (1/q) \sin \left( \mu + \mu_{n} \right) p_{n} \\
p_{n+1} & = & \left[ U_{n} \cos \left( \mu + \mu_{n} \right) -
q \sin \left(\mu + \mu_{n} \right) \right] x_{n} \\
& + & \cos \left( \mu + \mu_{n} \right) p_{n} \\
\end{array}
\label{hammap1}
\end{equation}
where $\mu=qa$ and $\mu_{n}=q\Delta_{n}$, and the values $x_{n}$ and $p_{n}$
refer to the instant before the $n-$th kick.

The evolution of the dynamical map~(\ref{hammap1}) can be analyzed
as follows. First, we make a weak-disorder expansion of
Eq.~(\ref{hammap1}), keeping only first- and second-order terms.
The expansion is straightforward; the resulting equations, however,
are lengthy and we omit them here.
As a second step, we perform a canonical transformation $(x_{n}, p_{n})
\rightarrow (X_{n}, P_{n})$, such that the unperturbed motion reduces to
a simple rotation in the phase space of the new variables~\cite{Izr01}.
Such a trick allows one to eliminate the effect of the periodic kicks
with constant amplitudes $U$. This can be done with the use of the
canonical transformation,
\begin{equation}
\begin{array}{ccl}
x_{n} & = & \alpha \cos (\mu/2) X_{n} +
(q \alpha)^{-1} \sin (\mu/2) P_{n} \\
p_{n} & = & - q \alpha \sin (\mu/2) X_{n} +
\alpha^{-1} \cos (\mu/2) P_{n}
\end{array}
\label{canonic}
\end{equation}
where the parameter $\alpha$ is defined by the relation
\begin{displaymath}
\alpha^{4} = \frac{1}{q^{2}}
\frac{\sin \mu  - \frac{U}{2q} \left( \cos \mu - 1 \right)}
{\sin \mu  - \frac{U}{2q} \left( \cos \mu + 1 \right)} .
\end{displaymath}
Note that, due to the transformation~(\ref{canonic}), the new
variables $X_{n}$ and $P_{n}$ have the same dimension.

In the absence of disorder, i.e., for $u_{n} = 0$ and $\Delta_{n} = 0$,
the rotation angle $\gamma$ between successive kicks is determined by the
relation,
\begin{eqnarray}
\cos \gamma = \cos \mu + \frac{U}{2q} \sin \mu & &
\mbox{ with } \gamma=ka.
\label{rotangle}
\end{eqnarray}
In terms of the Kronig-Penney model, $k$ is the Bloch wavevector,
$\gamma$ is the phase shift of the wavefunction within the lattice
step $a$, and Eq.~(\ref{rotangle}) defines the band structure of the
energy spectrum.

It should be pointed out that the transformation~(\ref{canonic}) is
well-defined for all values of the rotation angle $\gamma$ other
than $\gamma = 0$ and $\gamma = \pm \pi$ for which $\alpha$ either
vanishes or diverges. In other words, our approach fails at the
center and at the edges of the first Brillouin zone, i.e., at the
edges of the allowed energy bands of the KP model. However, the
approach works well in every neighborhood of these critical points.

To proceed further, it is useful to pass to the action-angle
variables $(J_{n}, \theta_{n})$, with the transformation
\begin{displaymath}
\begin{array}{ccc}
X_{n} = \sqrt{2 J_{n}} \sin \theta_{n}, & & P_{n} = \sqrt{2 J_{n}}
\cos \theta_{n}
\end{array}
\end{displaymath}
and to represent the Hamiltonian map~(\ref{hammap1}) in terms of the
new variables. Leaving aside mathematical details, we give here the
final expression,
\begin{equation}
\begin{array}{ccl}
J_{n+1} & = & D_{n}^{2} J_{n} \\
\theta_{n+1} & = & \theta_{n} + \gamma - \frac{1}{2}
\left[ 1 - \cos \left( 2 \theta_{n} + \gamma \right) \right] \tilde{u}_{n} \\
& + & \frac{1}{2}
\left[ \upsilon - \cos \left( 2 \theta_{n} + 2 \gamma \right) \right]
\tilde{\Delta}_{n} \\
\end{array}
\label{hammap5}
\end{equation}
where
\begin{equation}
\begin{array}{ccl}
D_{n}^{2} & = & 1 + \sin \left( 2 \theta_{n} + \gamma \right) \tilde{u}_{n} -
\sin \left( 2 \theta_{n} + 2 \gamma \right) \tilde{\Delta}_{n} \\
& + & \frac{1}{2}
\left[ 1 - \upsilon \cos \left( 2 \theta_{n} + 2 \gamma \right) \right]
\tilde{\Delta}_{n}^{2} \\
& + & \frac{1}{2}
\left[ 1 - \cos \left( 2 \theta_{n} + \gamma \right) \right]
\tilde{u}_{n}^{2} \\
& - & \left[ \cos \gamma  - \cos \left( 2 \theta_{n} + 2 \gamma \right) \right]
\tilde{u}_n \tilde{\Delta}_{n}.
\end{array}
\label{dnsquared}
\end{equation}
Here, $\upsilon = \left[ q \alpha^{2} + (q \alpha^{2})^{-1} \right]
q \sin \gamma/U$, and we have introduced the rescaled random variables,
\begin{eqnarray*}
\tilde{u}_{n} = \frac{\sin\mu}{q\sin\gamma}u_{n} & \mbox{and} &
\tilde{\Delta}_{n}=\frac{U}{\sin\gamma}\Delta_{n}.
\end{eqnarray*}
In Eqs.~(\ref{hammap5}) and~(\ref{dnsquared}) we have kept only the
terms of the weak-disorder expansion which are necessary to compute
the localization length within the second-order approximation. We
remark that the angle variable evolves independently of the action
variable.

The inverse localization length for the KP model~(\ref{kpmodel}) can
be computed as the Lyapunov exponent $\lambda$,
\begin{displaymath}
l^{-1}_{\rm loc} = \lambda = \lim_{N \rightarrow \infty}
\frac{1}{Na} \sum_{n=1}^{N} \ln
\left|\frac{\psi_{n+1}}{\psi_{n}}\right| = \Big \langle \frac{1}{a}
\ln \left|\frac{\psi_{n+1}}{\psi_{n}}\right| \Big \rangle
\end{displaymath}
which, in terms of the dynamical map~(\ref{hammap5}), can be written
as \cite{Izr98},
\begin{equation}
\lambda = \Big \langle \frac{1}{2a} \ln \left( \frac{J_{n+1}}{J_{n}}
\right) \Big \rangle  = \frac{1}{2a} \langle \ln D_{n}^{2} \rangle .
\label{lyap1}
\end{equation}
By expanding the logarithm of $D_{n}^2$, one gets
\begin{equation}
\begin{array}{ccl}
\lambda & = & \frac{1}{2a} \Big \langle \left\{
\sin \left( 2 \theta_{n} + \gamma \right) \tilde{u}_{n} -
\sin \left( 2 \theta_{n} + 2 \gamma \right) \tilde{\Delta}_{n}
\right. \\
& + & \frac{1}{4}\left[ 1 -
2 \upsilon \cos\left( 2 \theta_{n} + 2 \gamma \right)+
\cos \left( 4 \theta_{n} + 4\gamma \right) \right] \tilde{\Delta}_{n}^{2} \\
& + & \frac{1}{4} \left[ 1 - 2 \cos \left( 2 \theta_{n} + \gamma \right) +
\cos \left( 4 \theta_{n} + 2 \gamma \right) \right] \tilde{u}_{n}^{2} \\
& - & \left. \frac{1}{2}
\left[ \cos \gamma - 2 \cos \left( 2 \theta_{n} + 2 \gamma \right) +
\cos \left( 4 \theta_{n} + 3\gamma\right) \right]
\tilde{u}_{n} \tilde{\Delta}_{n} \right\} \Big \rangle . \\
\end{array}
\label{logd}
\end{equation}

Now, in order to obtain the Lyapunov exponent $\lambda$, we have to
perform the average over the phase $\theta_n$ and the random variables
$u_n$ and $\Delta_n$. To the second order of perturbation theory,
one can neglect the correlations between $\theta_{n}$ and the quadratic
terms $\tilde{u}_n^{2}$, $\tilde{\Delta}_{n}^{2}$, and
$\tilde{u}_{n} \tilde{\Delta}_{n}$. Hence for the summands in
Eq.~(\ref{logd}) which contain these quadratic terms, one can compute
separately the averages over $\theta_{n}$ and over the random variables
$u_n$ and $\Delta_n$.

In analogy with the Anderson model (see details and references
in~\cite{Izr98}), it can be shown that for our purposes it is
safe to assume that the invariant measure of the phase is a flat
distribution, $\rho(\theta)=1/(2\pi)$.
The assumption holds for all values of $\gamma$, except for
$\gamma=\pm \pi/2$ where a small modulation of the invariant measure
results in an anomaly for the localization length.
The situation with these values of $\gamma$ is similar to that known for
the standard 1D Anderson model at the center of the energy band, and the
correct expression for $\lambda$ can be obtained following the approach
of~\cite{Izr98}.

It should be stressed that weak modulations of $\rho(\theta)$ arise
also for other ``resonant'' values, $\gamma=m\pi/r$, with $m$
and $r$ integers prime with each other and $r > 2$.
However, these modulations do not influence the value of $\lambda$,
because the expression to be averaged in Eq.~(\ref{logd}) has no
harmonics higher than $4 \theta$.
Thus, our further analysis is valid for all values of $\gamma$ except
$\gamma=0$ and $\gamma = \pm \pi$ (i.e., the edges of the energy bands)
and $\gamma=\pm \pi/2$.

After averaging, the expression for the Lyapunov exponent takes the
form,
\begin{equation}
\begin{array}{ccl}
\lambda & = & \displaystyle
\frac{1}{8a}
\left[ \langle \tilde{u}_{n}^{2} \rangle +
\langle \tilde{\Delta}_{n}^{2} \rangle -
2 \langle \tilde{u}_{n} \tilde{\Delta}_{n} \rangle \cos \gamma \right] \\
& + & \displaystyle
\frac{1}{2a} \langle \tilde{u}_{n} \sin(2 \theta_n + \gamma) \rangle -
\frac{1}{2a} \langle \tilde{\Delta}_{n}
\sin \left( 2 \theta_{n} + 2 \gamma \right) \rangle. \\
\end{array}
\label{lyap2}
\end{equation}
In order to compute the noise-angle correlators in Eq.~(\ref{lyap2}),
we generalize the method used in~\cite{Izr99}. Specifically, we
introduce the correlators
$r_{k} = \langle \tilde{u}_{n} \exp \left( i 2 \theta_{n-k} \right) \rangle$
and
$s_{k}=\langle \tilde{\Delta}_{n} \exp \left( i 2 \theta_{n-k} \right)\rangle$.
Both correlators satisfy recursive relations that can be obtained by
substituting the angular map of Eq.~(\ref{hammap5}) into the definitions of
$r_{k-1}$ and $s_{k-1}$. The recursive relations allow one to obtain the
correlators $r_{0}$ and $s_{0}$, whose imaginary parts represent the
noise-angle correlators in Eq.~(\ref{lyap2}). As a result, we
arrive at the final expression for the Lyapunov exponent,
\begin{equation}
\begin{array}{ccl}
\lambda & = & \displaystyle
\frac{1}{8a}\left[ \langle \tilde{u}_{n}^{2} \rangle W_{1} +
\langle \tilde{\Delta}_{n}^{2} \rangle W_{2} -
2 \langle \tilde{u}_{n} \tilde{\Delta}_{n} \rangle \cos \gamma W_{3} \right] \\
& = & \displaystyle
\frac{\sin^{2}\mu}{8 aq^{2}\sin^{2}\gamma}
\langle u_{n}^{2} \rangle W_{1} +
\frac{U^{2}}{8 a \sin^{2}\gamma} \langle \Delta_{n}^{2} \rangle W_{2} \\
& - & \displaystyle
\frac{1}{4a} \frac{U \sin \mu}{q \sin^{2} \gamma} \cos \gamma \;
\langle u_{n} \Delta_{n} \rangle W_{3}
\end{array}
\label{invloc}
\end{equation}
where the functions
\begin{equation}
W_{i} \equiv W_{i} \left( 2 \gamma \right) = 1 + 2 \sum_{k=1}^{\infty}
\chi_{i}(k) \cos(2\gamma k) \;\;\;\; (i=1,2,3) \label{ftcorr}
\end{equation}
are the $2\gamma-$harmonics of the Fourier transform of the binary
correlators $\chi_{i} (k)$, see Eq.~(\ref{bincor}). We should stress
that Eq.~(\ref{invloc}) has been derived without invoking the Born
approximation, $E \gg U$; it describes the tunneling regime for $E < U$
as well as the scattering one, $E > U$. The only constraint is the
weakness of both types of disorder, $\langle u_{n}^{2} \rangle \ll U^{2}$
and $q^{2} \langle \Delta_{n}^{2} \rangle \ll a^{2}$.

Let us first discuss the structure of expression~(\ref{invloc}) for
the case in which there are no correlations between $u_{n}$ and
$\Delta_{n}$. It is quite instructive that for weak scattering,
$U \ll q$, (therefore, $\gamma \approx \mu$) with purely compositional
disorder, (i.e., $\Delta_{n} = 0$), the Lyapunov exponent takes the form
$\lambda \approx \frac{ \langle u_{n}^{2} \rangle}{8 aq^{2}} W_{1}$,
(see~\cite{Kro02}). This is equivalent to the well-known result for weak
scattering in continuous 1D potentials,
$\lambda= \frac{\langle V^{2}\rangle}{8q^2}W(2q)$,
where $\langle V^{2} \rangle$ is the variance of the random potential
and $W(2q)$ is the $2q-$component of the power spectrum of the potential.

In the other limit case of structural disorder (i.e., $u_{n}=0$), the
expression for the Lyapunov exponent,
$\lambda = \frac{U^{2}}{8 a \sin^{2}\gamma}
\langle \Delta_{n}^{2} \rangle W_{2}$,
was obtained in~\cite{Izr01}. One can see that it is similar to the
expression for the tight-binding Anderson model with diagonal disorder,
$\lambda =\frac{\langle\varepsilon^{2} \rangle}{8 a\sin^{2}\mu} W (2\mu)$,
where $\langle \varepsilon^{2} \rangle$ is the variance of the random
site-potential, $\mu$ is the phase shift of the wavefunction between two
sites -related to the energy by the dispersion relation $E = 2 \cos \mu$-,
and $W(2\mu)$ has the same meaning as $W_{1}$.

Expression~(\ref{invloc}) for the inverse localization
length allows one to estimate the relative importance of the
structural and compositional disorder for the transport properties of
a finite sample. As can be seen, spatial correlations of the variables
$u_{n}$ and $\Delta_{n}$ can enhance or suppress the localization length
in comparison with the case of uncorrelated disorder.

Specific long-range correlations can make the Fourier
transforms~(\ref{ftcorr}) vanish in prescribed energy windows, so that
the Lyapunov exponent~(\ref{invloc}) also vanishes in the same energy
intervals.
A method for the construction of random potentials with given binary
correlators $\chi_{i} (k)$ was described and tested in~\cite{Izr99,Izr05}.
Following the same approach, one can use formula~(\ref{invloc}) as
a starting point for the fabrication of devices with prescribed anomalous
transport characteristics.

It is interesting to relate the properties of the KP model with known
results for 1D tight-binding models with both diagonal and off-diagonal
disorder. It can be shown that, after eliminating the momenta $p_{n}$ from
the map~(\ref{hammap1}), one obtains the equation
\begin{equation}
\begin{array}{l}
\left[ 1/\sin (\mu + \mu_{n}) \right] \psi_{n+1} +
\left[ 1/\sin (\mu + \mu_{n-1}) \right] \psi_{n-1} \\
= \left[ \cot ( \mu + \mu_{n} ) + \cot ( \mu + \mu_{n-1} ) +
\frac{1}{q} \left( U + u_{n} \right) \right] \psi_{n}
\end{array}
\label{tightbind}
\end{equation}
where $\psi_{n} \equiv x_{n}$. For weak disorder one can expand the
coefficients of Eq.~(\ref{tightbind}) in powers of $q\Delta_{n}$ and get
\begin{equation}
\begin{array}{cl}
& \displaystyle
\left[ 1 + q^{2}\left( 1/2 + \cot^{2}\mu \right)
\langle \Delta_{n}^{2} \rangle -q \Delta_{n} \cot \mu    \right] \psi_{n+1} \\
+ & \displaystyle
\left[ 1 + q^{2}\left( 1/2 + \cot^{2}\mu \right)
\langle \Delta_{n}^{2} \rangle - q \Delta_{n-1}\cot \mu  \right] \psi_{n-1} \\
= & \left[ 2 \cos \mu +  U\frac{\sin\mu}{q} + 2q^{2}
\frac{\cos\mu}{\sin^{2}\mu}  \langle \Delta_{n}^{2} \rangle
\right. \\
- & \left. \frac{q}{\sin\mu} \left( \Delta_{n} + \Delta_{n-1} \right)
+ \frac{\sin\mu}{q} u_n \right] \psi_{n} .
\end{array}
\label{tibi}
\end{equation}
One can see that the right-hand side of this relation vanishes under
the conditions
\begin{eqnarray}
U & = & -\frac{2q \cos\mu}{\sin\mu}
\left[ 1 + \frac{q^{2}}{\sin^{2}\mu} \langle \Delta_{n}^{2} \rangle \right],
\label{con1} \\
u_{n} & = &
\frac{q^{2}}{\sin^{2}\mu} \left( \Delta_{n} + \Delta_{n-1} \right) .
\label{con2}
\end{eqnarray}
The first condition determines the energy as a function of the
mean field $U$ and of the variance $\langle \Delta_{n}^{2} \rangle$.
The second condition establishes non-trivial correlations between the
compositional and structural disorders. Under these conditions
Eq.~(\ref{tibi}) reduces to
\begin{equation}
\begin{array}{l}
\displaystyle
\left[ 1 + q^{2}\left( 1/2 + \cot^{2}\mu \right)
\langle \Delta_{n}^{2} \rangle - q \Delta_{n}\cot \mu   \right] \psi_{n+1} \\
\displaystyle
+ \left[ 1 + q^{2}\left( 1/2 + \cot^{2}\mu \right) \langle
\Delta_{n}^{2} \rangle -q \Delta_{n-1}\cot \mu   \right] \psi_{n-1}
= 0 \\
\end{array}
\label{specialkp}
\end{equation}
and this identity can be written as the Schr\"{o}dinger equation
\begin{equation}
\left( \gamma + \gamma_{n+1} \right) \psi_{n+1} +
\left( \gamma + \gamma_{n} \right) \psi_{n-1} = 0
\label{anomand}
\end{equation}
for the Anderson model with purely off-diagonal disorder at the
center of the energy band.

As is known~\cite{Che05}, the model~(\ref{anomand}) exhibits
anomalous localization. Specifically, the band-center electronic
state is localized but decays away from the localization
center $n_{0}$ as $\psi_{n} \sim \exp \left( - A \sqrt{|n - n_{0}|}
\right)$ where $A$ is some constant. Thus, the interplay between
compositional and structural disorder can give rise to anomalous
localization in the KP-model.

We have to note that the above conclusion about the anomalous
localization cannot be drawn directly from the general
formula~(\ref{invloc}) for two reasons. First, the zero-value of the
Lyapunov exponent leaves open the question of whether the
corresponding electronic state is extended or anomalously localized.
Second, the conditions~(\ref{con1}) and~(\ref{con2}) imply
that the Bloch vector $k$ lies in a neighborhood of the points $\pm
\pi/2a$, where our derivation of the Lyapunov exponent may be
invalid. Indeed, when conditions~(\ref{con1}) and~(\ref{con2}) are
met, the requirement of weak compositional disorder, $\langle
u_n^{2} \rangle \ll U^{2}$ , leads to $q^{2} \langle \Delta_{n}^{2}
\rangle \ll \sin^{2}\mu$ . Taking into account this inequality and
the relation~(\ref{rotangle}) with $U$ given by~(\ref{con1}), one
can see that the Bloch wavenumber is actually close to the resonant
values $\pm \pi/2a$.

In conclusion, we have derived the expression for the localization
length in the KP model for the general case when both the amplitudes and
the spacings of the $\delta$-barriers are random variables. Our
consideration takes into account correlations of each type of disorder
with itself, as well as correlations between the two disorders.
The obtained expression depends on the binary correlators only and
thus opens the way to the construction of random potentials with
desired characteristics of the electron (or electromagnetic waves)
transmission through finite samples.
The most important application of expression~(\ref{invloc}) lies in its
use for the fabrication of devices with prescribed energy windows
with perfect transmission (or reflection) produced by long-range
correlations of the disorder, an effect recently  observed experimentally
in microwave waveguides~\cite{Kuh00,Kro02}.
We have also found that for specific correlations between the two kinds
of disorder, the Kronig-Penney model has anomalously localized
eigenstates, similarly to the Anderson model with off-diagonal
disorder.

The authors gratefully acknowledge the financial support of the Coecyt
grant CB0702201-X.

\end{document}